\documentclass[a4,journal]{IEEEtran}
\usepackage{afterpage}
\usepackage{stfloats}
\usepackage{float}
\usepackage{amsmath,amsfonts}
\usepackage{algorithmic}
\usepackage{array}
\usepackage[caption=false,font=normalsize,labelfont=rm,textfont=sf]{subfig}
\usepackage{amssymb}
\usepackage{textcomp}
\usepackage{stfloats}
\usepackage{url}
\usepackage{verbatim}
\usepackage{graphicx}
\usepackage{bm}
\usepackage{makecell}
\usepackage{colortbl}
\usepackage{color}
\usepackage[table,xcdraw]{xcolor}
\usepackage{nomencl}
\usepackage{setspace}
\usepackage{color}
\usepackage{tabularx}
\def\BibTeX{{\rm B\kern-.05em{\sc i\kern-.025em b}\kern-.08em
  T\kern-.1667em\lower.7ex\hbox{E}\kern-.125emX}}
\usepackage{balance}
\usepackage{booktabs, tabularx}
\usepackage[referable]{threeparttablex}
\usepackage{stfloats}
\usepackage{multirow}
\usepackage{float}
\usepackage{diagbox}
\usepackage{makecell}
\usepackage[ruled, linesnumbered]{algorithm2e}
\usepackage{cite}
\usepackage[colorlinks=true, linkcolor=blue, citecolor=blue, urlcolor=blue]{hyperref}

\usepackage{pifont}  

\newcommand{\cmark}{\ding{51}} 
\newcommand{\xmark}{\ding{55}} 
\usepackage{capt-of}
\begin{document}

\title{Gaussian Processes in Power Systems: Techniques, Applications, and Future Works}

\author{Bendong Tan,~\IEEEmembership{Member,~IEEE}, Tong Su,~\IEEEmembership{Student Member,~IEEE}, Yu Weng,~\IEEEmembership{Member,~IEEE}, Ketian Ye,~\IEEEmembership{Member,~IEEE}, Parikshit Pareek,~\IEEEmembership{Member,~IEEE}, Petr Vorobev,~\IEEEmembership{Member,~IEEE}, Hung Nguyen,~\IEEEmembership{Member,~IEEE}, Junbo Zhao,~\IEEEmembership{Senior Member,~IEEE}, Deepjyoti Deka,~\IEEEmembership{Senior Member,~IEEE} 
\thanks{B. Tan, T. Su and J. Zhao are with the Department of Electrical and Computer Engineering, University of Connecticut, Storrs, CT, USA. Y. Weng and D. Deka are with the MIT Energy Initiative, Cambridge, MA, USA. K. Ye is with Siemens PTI, Minnetonka, MN, USA. P. Pareek is with the Department of Electrical Engineering, Indian Institute of Technology Roorkee, Uttarakhand, India. P. Vorobev and H. Nguyen are with the School of Electrical and Electronic Engineering, Nanyang Technological University, Singapore. (email: junbo@uconn.edu)}
\thanks{}
}

\markboth{}%
{Shell \MakeLowercase{\textit{et al.}}: A Sample Article Using IEEEtran.cls for IEEE Journals}


\maketitle

\begin{abstract}
The increasing integration of renewable energy sources (RESs) and distributed energy resources (DERs) has significantly heightened operational complexity and uncertainty in modern power systems. Concurrently, the widespread deployment of smart meters, phasor measurement units (PMUs) and other sensors has generated vast spatiotemporal data streams, enabling advanced data-driven analytics and decision-making in grid operations. In this context, Gaussian processes (GPs) have emerged as a powerful probabilistic framework, offering uncertainty quantification, non-parametric modeling, and predictive capabilities to enhance power system analysis and control. This paper presents a comprehensive review of GP techniques and their applications in power system operation and control. GP applications are reviewed across three key domains: GP-based modeling, risk assessment, and optimization and control. These areas serve as representative examples of how GP can be utilized in power systems. Furthermore, critical challenges in GP applications are discussed, and potential research directions are outlined to facilitate future power system operations.

\end{abstract}
\vspace{-0.2cm}

\section{Introduction} 
The extensive reliance on fossil fuels has drastically elevated carbon emissions, presenting a substantial threat to global climate stability and sustainable development. As the principal medium for energy production and transmission, power systems are positioned at the forefront of the energy transition, playing an indispensable role in global carbon reduction efforts \cite{Lopion2018review}. This transition has catalyzed the widespread adoption of renewable energy sources (RESs), distributed energy resources (DERs), and flexible loads, such as electric vehicles (EVs) \cite{cao2020reinforcement}. While these components contribute to decarbonization, they simultaneously introduce considerable uncertainties and volatilities, posing challenges to the secure operation of power systems. Furthermore, as these elements are primarily integrated into the grid through power electronic devices, the overall system complexity is substantially increased, complicating both structural and parametric aspects of system modeling. Consequently, physics-based power system analyses may become either time-intensive or imprecise in the face of such uncertainties and complexities.

The proliferation of advanced sensing technologies, such as phasor measurement units (PMUs), smart meters, Internet of Things-related sensors, etc, has resulted in power systems generating more comprehensive and valuable data than ever before. This evolution is propelling power systems into a new era of digitalization and intelligence, offering a promising avenue for addressing the aforementioned challenges. As a result, there is a growing demand for data-driven analytical methods that not only extract critical steady-state and dynamic insights from data, but also improve the accuracy of physical system modeling.

A variety of machine learning (ML) techniques have been introduced for power system analysis, including shallow learning methods such as support vector machines (SVM) \cite{Hearst1998Support}, as well as newer deep learning models like Transformer \cite{NEURIPS2021_854d9fca}. These approaches offer the advantage of being model-free and capable of generalizing to new power system scenarios, providing a degree of adaptability. However, learning models solely on data makes them prone to overfitting or incorporating unwanted biases, especially when the dataset is small \cite{jakkala2021deep}. To address these limitations, Bayesian methods have been developed to integrate prior knowledge into model learning. Of particular note is the framework of GP, an important Bayesian learning paradigm, that generalizes multivariate Gaussian distributions by placing distributions over functions \cite{bishop2006pattern,williams2006gaussian}. This enables GP to uncover hidden functional relationships in power system tasks when sufficient observations are collected. GP offers several key advantages that make it feasible for use in the analysis of critical infrastructure such as power systems, where safety is a priority.
\begin{itemize}
\item \textit{Flexible expressivity.} The nonparametric nature of GP allows it to model highly complex functions \cite{tazi2023beyond}, making it ideal for capturing the intricate dynamics of modern power systems, which may include complex devices with proprietary internal control structures, such as inverter-based resources (IBRs). 
\item \textit{Well-calibrated uncertainty quantification.} A key advantage of GP is its ability to provide closed-form uncertainty estimates, accounting for both data noise and discrepancies from new scenarios that differ from the training data distribution \cite{tazi2023beyond}. This is essential for use within safety-critical systems such as power grids. 
\item \textit{Efficient with small datasets.} By using Bayesian learning principles \cite{williams2006gaussian} and with only a few hyperparameters to optimize, GP models can generalize effectively even with limited data \cite{snelson2007local}, which is particularly beneficial for applications where training data is scarce. Additionally, GP training can be interlaced with active learning \cite{pareek2024fast} to overcome data scarcity if generating data is expensive and time-consuming.
\end{itemize}

Due to these advantages, GP has garnered increasing attention in the field of power systems in recent years, particularly for applications such as forecasting, risk analysis, and decision-making. This article presents the fundamentals of GP modeling and then explores its use cases in power grid analysis. To the best of our knowledge, this paper is the first to provide a comprehensive survey of GP applications in power systems. In addition, we discuss the key challenges and future directions for the use of GP in this field, as well as their integration with other data-driven methods. The main contributions of this paper are summarized as follows:
\begin{itemize}
\item This paper presents a comprehensive and structured overview of the fundamental concepts of GP, along with an overview of the supporting software ecosystem.
\item This paper highlights key applications of GP in power systems, including load and renewable energy forecasting, power system static and dynamic modeling, risk assessment, and decision-making. The application procedures and prior work are discussed in detail, demonstrating how GP effectively addresses challenges in these domains.
\item This paper investigates the major challenges in applying GP to power systems, such as scalability and robustness, and discusses applications where GP may have greater applicability. Lessons learned are discussed and finally, future research directions, particularly in integrating GP with other advanced data-driven methods, are outlined.
\end{itemize}

The remainder of this paper is organized as follows. Section \ref{sec:basic} offers an overview of the fundamental principles and ecosystem of GP. Section \ref{sec:application} provides a comprehensive review of GP-based learning in power system applications: forecasting, static and dynamic estimation and solutions to optimization problems. Next, we show how such learned models can be used for static and dynamic risk assessment in power grids in Section \ref{sec:risk}, and inside optimization and control modules in Section \ref{sec:optandcontrol}. Finally, Section \ref{sec:challenge} discusses the challenges and proposes future research directions for the application of GP in power systems. Section \ref{sec:conclusion} concludes the paper.

\begin{figure*}
  \centering
  \includegraphics[width=0.9\linewidth]{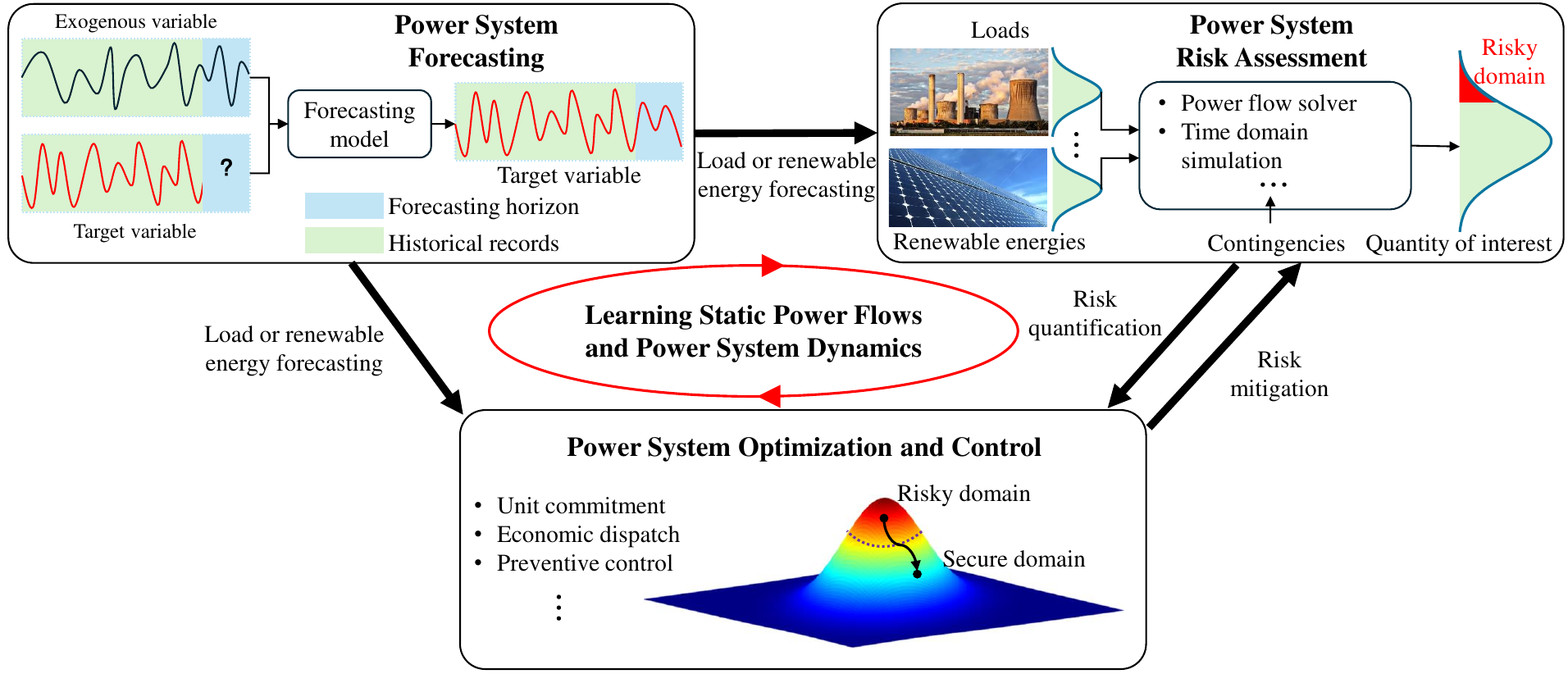}
  \caption{Illustration of GP applications across various domains of power systems. GP-based learning techniques have been used to support load and renewable energy forecasting, static and dynamic modeling of power flows, and are integrated into power system optimization and control. These methods further contribute to power system risk assessment through improved forecasting, risk quantification, and mitigation strategies, enhancing secure and reliable grid operation. See Table \ref{forecasting_summary} and Table \ref{Table_Optimization} for specific works related to specific problems.}
  \label{fig:GP application framework}
\end{figure*}

\section{Preliminaries of Gaussian Process}\label{sec:basic}
\subsection{Gaussian Process Fundamentals}
This section presents the basic details of GP from a functional space perspective. GP is a non-parametric and probabilistic modeling method. Consider a latent function $f(\cdot)$ that describes a data point $(x_i, y_i)$ as follows:
\begin{equation}
  y_i = f(x_i) + \varepsilon_i,
\end{equation}
where $y_i$ is the observation with respect to the input $x_i$ of $i$-th sample, and $\varepsilon_i$ is a sample drawn from a Gaussian noise of zero mean and variance $\sigma^2_\varepsilon$. 

Let, the function values $\boldsymbol{f}(x) = [f(x_1), f(x_2), \dots f(x_N)]^\top$ for $N$ inputs $x =(x_1, x_2, \dots x_k)$ be random values that are jointly Gaussian, i.e.,
\begin{align}
    \boldsymbol{f}(x) \sim \mathcal{N}(\bm{m}(x), K(x, x')),
\end{align}
where $m(x)$ and $K(x, x)$ are the parameterized mean vector and covariance matrix, respectively, that depend on $x$. For example, for two input points, Gaussianity implies
\begin{align}\label{eq:gp_f}
   \begin{bmatrix}
    f(x_1) \\ f(x_2)
  \end{bmatrix} \sim \mathcal{N}\bigg ( \underbrace{\begin{bmatrix}
    m(x_1) \\ m(x_2)
  \end{bmatrix}}_{\texttt{Mean}}, \underbrace{\begin{bmatrix}
    k(x_1, x_1) & k(x_1, x_2) \\ k(x_2, x_1) & k(x_2, x_2)
  \end{bmatrix}}_{\texttt{Covariance Matrix}} \bigg ).
\end{align}
By definition \cite{williams2006gaussian, murphy2022probabilistic},
\begin{quote}
  \textit{A Gaussian Process is a (potentially infinite) collection of random variables such that any finite subset of it has a joint multivariate Gaussian distribution.}
\end{quote}
$f$ is termed a GP, denoted as $f \sim \mathcal{GP}(m, k)$. The distribution $\mathbb{P}(y)$ of the observation $y$ at inputs $x$ is given by marginalizing out $f$ to give 
\begin{align}
  \mathbb{P}(y) = \int \mathbb{P}(y|f)d\mathbb{P}(f) \sim \mathcal{N}(m(x), K(x, x')+\sigma_\epsilon^2\mathbb{I}).
\end{align}

The covariance matrix $K$ is composed of entries $K_{ij} = k(x_i, x_j)$ that measure the covariance of $f$ at two inputs $x_1$ and $x_2$. Both the mean and covariance function can be selected to encode prior knowledge about the shape of $f(\cdot)$. The Gaussianity of $f(x)$ and consequently of $y$ has a particularly attractive property, based on the efficient computation of conditional probability for Gaussian random variables. Consider that we want to determine the value of the function at a new input point $x^\star$. From the joint Gaussianity property, we know that observed values $y$ at training points and the new function value $f^\star$ is given by:
\begin{align}
  \begin{bmatrix}
    \bm{y} \\ f^\star
  \end{bmatrix} \sim \mathcal{N}\bigg( \begin{bmatrix}
    \bm{m}(x) \\ m(x^\star)
  \end{bmatrix}, \begin{bmatrix}
    K(x,x) + \sigma^2_\varepsilon \mathbb{I} & K(x,x^\star) \\ K(x^\star,x) & K(x^\star,x^\star)
  \end{bmatrix} \bigg)\label{eq:noise}
\end{align}
where, $x$ is the vector of all input data points (training dataset) that we want to use for conditioning, $\mathbb{I}$ denotes a identity matrix, $K(x,x)$ is the covariance matrix among training points, $K(x,x^\star)$ is the covariance between training points and a new input point, and $K(x^\star,x^\star)$ is the self-covariance of the new input point. As conditioning on training data implies that we want to know the value of $f^\star$ given $y$, we can use the Gaussian distribution conditioning result to obtain the distribution of $f^\star$, as given in \eqref{eq:fstarnoise}.

\begin{figure*}[htb]
\begin{align}
  f^\star|\bm{y} \sim& \mathcal{N}\Big( m(x^\star) + K(x^\star,x)\big[K(x,x) + \sigma^2_\varepsilon \mathbb{I}\big]^{-1}(\bm{y} - \bm{m}(x)),
  K(x^\star,x^\star) - K(x^\star,x)\big[K(x,x)+ \sigma^2_\varepsilon \mathbb{I}\big]^{-1}K(x,x^\star) \Big). \label{eq:fstarnoise}
\end{align}
\end{figure*}
The mean and covariance functions are parameterized to adapt the GP model based on the training data. These parameters are referred to as hyperparameters. Table \ref{tab:hyper} shows a few commonly used covariance/kernel functions and their associated hyperparameters. A standard strategy is often to use a mean function of zero. To learn these hyperparameters from the training data, one can use maximum log-likelihood on the marginal distribution $\mathbb{P}(y)$ of observations $y$ at inputs $x$, as\footnote{See chapter 2 of \cite{williams2006gaussian} for derivation.}:
\begin{align}\label{eq:mll}
   -\frac{1}{2}\big [\bm{y}^\top[K+\sigma^2_\varepsilon \mathbb{I}] \bm{y} + \log |K+\sigma^2_\varepsilon \mathbb{I}| + N \log 2\pi \big ].
\end{align}

\begin{figure}
  \centering
  \includegraphics[width=\linewidth]{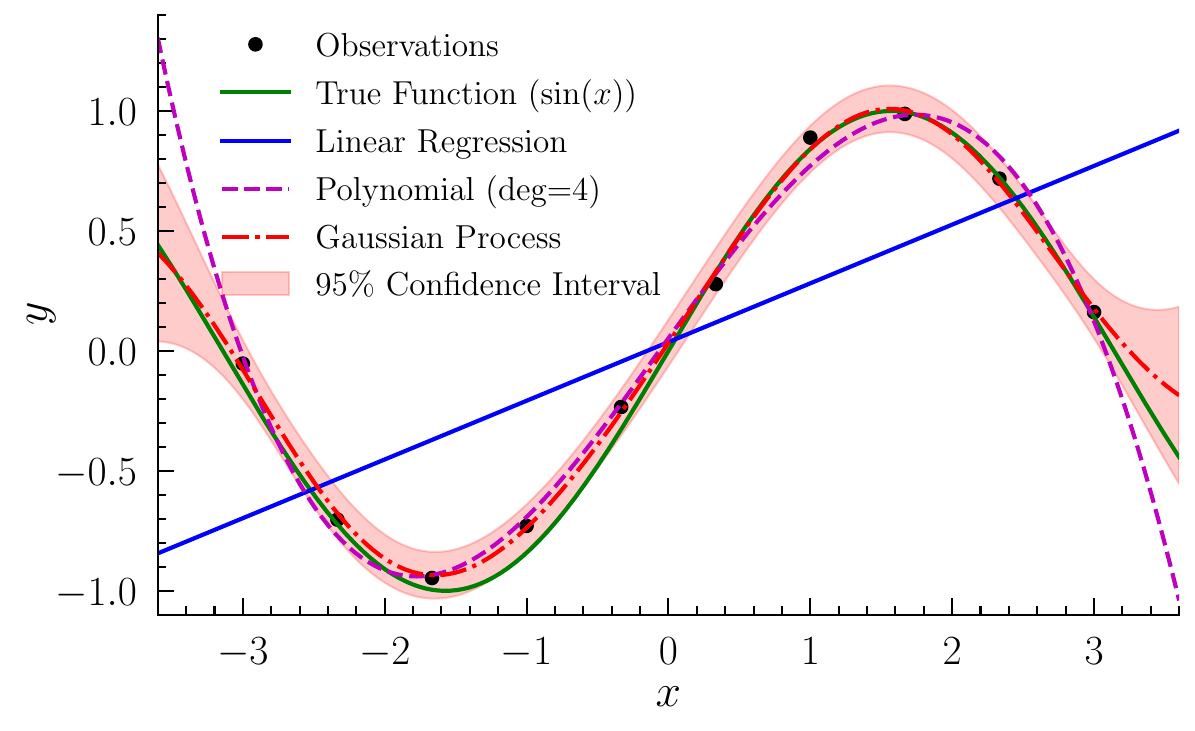}
  \caption{Comparison of linear, polynomial, and GP regressions (using Square Exponential kernel with \(\nu^2=1\)) on \( y = \sin(x) + 0.05 \cdot \varepsilon \), where \( \varepsilon \sim \mathcal{N}(0, 1) \). Importantly, confidence interval information is only available with GP regression.}
  \label{fig:GP_REGRESSION}
\end{figure}

Fig.~\ref{fig:GP_REGRESSION} shows the prediction comparison among linear, polynomial, and GP regressions on the sine function with added noise. All models are trained using only 10 observations. While the linear and polynomial models provide point estimates with limited flexibility, the GP regression not only captures the nonlinear pattern of the data more accurately but also quantifies predictive uncertainty through variance estimates, providing confidence intervals around its predictions. This demonstrates the advantages of GP in modeling complex nonlinear relationships and evaluating the reliability of its outputs.

\begin{table}[t]
  \centering
     \footnotesize
    \centering
    \captionof{table}{Standard Kernels for GP}
    \label{tab:hyper}
    \begin{threeparttable}
      \begin{tabular}{>{\centering\arraybackslash}p{3cm} >{\centering\arraybackslash}p{5cm}} 
        \toprule
        \midrule
        \textbf{Kernel Name} & \textbf{Function} \(k(x, x')\) \\
        \midrule
        Linear & 
        \(\nu^2 (x \cdot x') \) \\
        
        Polynomial & 
        \((\alpha (x \cdot x') + c)^d \) \\
        
        Square Exponential & 
        \( \nu^2 \exp\left(-\frac{\|x - x'\|^2}{2l^2}\right) \) \\
        
        Matérn & 
        \(\frac{2^{1-\nu}}{\Gamma(\nu)} \left( \frac{\sqrt{2\nu}\|x - x'\|}{l} \right)^\nu K_\nu \left( \frac{\sqrt{2\nu}\|x - x'\|}{l} \right) \) \\\midrule
        \bottomrule
      \end{tabular}
      \begin{tablenotes}
        \small
        \item \textbf{Hyperparameters}-- \(\nu^2\): variance; \(\alpha\): scale; \(c\): offset; \(d\): degree; \(l\): length scale; \(\nu\): smoothness. See Chapter 4 of \cite{williams2006gaussian} for more details.
      \end{tablenotes}
    \end{threeparttable}
\end{table}

\begin{table*}[t]
\footnotesize
  \centering
  \caption{Comprehensive Comparison of Selected GP Software Packages}
  \label{tab:gp_packages}
  \renewcommand{\arraystretch}{1.2}
  \begin{threeparttable}
  \begin{tabular}{c c >{\centering\arraybackslash}p{1cm} >{\centering\arraybackslash}p{1.5cm} >{\centering\arraybackslash}p{1.6cm} | c c >{\centering\arraybackslash}p{1cm} >{\centering\arraybackslash}p{1.5cm} >{\centering\arraybackslash}p{1.5cm}}
    \toprule
    \midrule
    \thead{\textbf{Package}} & \thead{\textbf{Language}} & \thead{\textbf{GPU} \\ \textbf{Support}} & \thead{\textbf{Deep} \\ \textbf{Learning}} & \textbf{Maintenance} 
    & \thead{\textbf{Package}} & \thead{\textbf{Language}} & \thead{\textbf{GPU} \\ \textbf{Support}} & \thead{\textbf{Deep} \\ \textbf{Learning}} & \textbf{Maintenance} \\
    \midrule
    GPy        & Python & \cmark & \cmark (Basic) & Active  & scikit-learn GP  & Python & \xmark & \xmark     & Active  \\
    GPflow       & Python & \cmark & \cmark (Basic) & Active  & PyMC3       & Python & \xmark & \xmark     & Active  \\
    GPyTorch      & Python & \cmark & \cmark (Full)  & Active  & BoTorch      & Python & \cmark & \cmark (Full)  & Active  \\
    Pyro        & Python & \cmark & \cmark (Full)  & Active  & GPML          &  MATLAB   &  \xmark   &  \xmark       &    \xmark \\\midrule
    \bottomrule
  \end{tabular}

  \begin{tablenotes}
   \small
   \item \textbf{Deep Learning}: Full = Deep kernel learning (DKL) \cite{wilson2015kernel,wilson2016deep} + Deep GP; Basic = Simple neural network integration.
   \item \textbf{Maintenance}: Active (Regular updates); Inactive (No updates in past 2 years).
  \end{tablenotes}
  \end{threeparttable}
\end{table*}

\textbf{Relation to non-linear Bayesian regression:} The expressions of GP, including those between training and new data, can be derived by considering a Bayesian non-linear regression model $y = \Phi(x)\beta + \epsilon$, for a non-linear feature vector $\Phi(x)$, Gaussian noise $\epsilon$ and parameter vector $\beta$ that follows a Gaussian prior distribution. If $\Phi(x)$ is defined by a covariance matrix $K(x,x)$, similar expressions for posterior distribution of $y$ and $\Phi(x^*)$ given training data $x$ can be derived by a MAP (Maximum APosteriori) estimator for $\beta$ using properties of Gaussian random variables and standard identities for matrix manipulation \cite{williams2006gaussian}.

\subsection{Gaussian Process Ecosystem}
A wide range of GP libraries have been extensively developed across various programming languages, including \texttt{Python}, \texttt{MATLAB}, \texttt{Julia}, and \texttt{R}. In \texttt{MATLAB}, the \texttt{GPML} package associated with the GP book \cite{williams2006gaussian} provides an excellent starting point for learning GP workings. GP libraries in the Python ecosystem, such as \texttt{GPflow} \cite{matthews2017gpflow} and \texttt{GPyTorch} \cite{gardner2018gpytorch}, offer significant advantages in deep learning integration and GPU acceleration, making them particularly suitable for uncertainty analysis and large-scale spatiotemporal data modeling. Also, \texttt{BoTorch} \cite{balandat2020botorch} provides necessary support for Bayesian Optimization using GP, along with large-scale GP regression. Further, \texttt{Julia} has several packages for working with GP, including \texttt{KernelFunctions.jl}, \texttt{GPLikelihoods.jl}, \texttt{AbstractGPs.jl}, and \texttt{ApproximateGPs.jl}. A few of these packages have been summarized in Table \ref{tab:gp_packages}.

\section{Gaussian Process based Modeling in Power System}\label{sec:application}
In this section, we present four specific power system use-cases where GP has been directly used for modeling or learning. 

\subsection{Forecasting: Loads and Renewable Energy Generation} 
The objective of load and renewable energy forecasting is to predict future power consumption and generation, both of which are essential for the efficient operation and strategic planning of power systems \cite{yang2018power}. Based on the time horizon, power system forecasting tasks can be divided into four categories: ultra-short-term, short-term, medium-term, and long-term forecasting \cite{zhang2014review, Ghasempour2023Short}. The application scenarios differ across these time scales, as summarized in Table \ref{forecasting}, which provides a comparison of forecasting tasks for each horizon. Since ultra-short-term and short-term forecasting are important for energy management and real-time operations of power systems, the majority of research has concentrated on these two forecasting horizons \cite{zhang2014review}.

\textbf{Problem formulation:} The forecasting model is developed to capture the relationship between historical data and future outcomes, and can be formulated as:
\begin{equation}
  \hat{y}_{t+k|t}= f(\boldsymbol{x}_t|\boldsymbol{\theta}_\text{forecast}),
  \label{forecasting_model}
\end{equation}
where $\hat{y}_{t+k|t}$ represents the targeted prediction at time $t+k$; $f$ denotes the forecasting function; $\boldsymbol{x}_t$ is the historical information vector, which may include indicators such as calendar time and weather data relevant to $\hat{y}_{t+k|t}$; and $\boldsymbol{\theta}$ denotes the hypeparameters of the forecasting model. Note that the output for solar can be either solar photovoltaic (PV) generation power or solar irradiance, with the latter being converted into PV generation power using a conversion function \cite{zhang2014review}. Similarly, wind forecast output can be either wind power generation or wind speed, with the latter being converted into wind power generation using the following conversion function \cite{anaya2011wind}. The vector $\boldsymbol{x}_t$ can include historical records of the target variable $y$ and exogenous variables $\boldsymbol{z}$ as well as forecasts of $\boldsymbol{z}$.

Numerous machine learning methods have been used for forecasting load and renewable energy \cite{HOU2022Review}, including new Long Short-Term Memory Networks (LSTM) \cite{Kong2019Short} and temporal convolutional networks \cite{Zuo2021Short}. Renewables such as wind and solar depend on variable and chaotic weather conditions \cite{Sheraz2021survey}, while power loads are impacted by weather and additional factors such as calendar effects, socioeconomic conditions \cite{Fan2012Short}. As point forecasting methods may not adequately capture uncertainties, we discuss the use of GP-based forecasting in this area.

\textbf{GP-based forecasting:} Under the GP framework, the target variable values can be modeled as Gaussian-distributed realizations of $\hat{y}_{t}$ \cite{Li2024Residential}:
\begin{align}
y_{t+k|t} & \sim \mathcal{N}\left(\hat{y}_{t+k|t}, K(\boldsymbol{x}_{t}, \boldsymbol{x}_{t}')+\sigma_\epsilon^2\mathbb{I}\right), \\
\text{where}\quad \hat{y}_{t+k|t} & = m\left(\boldsymbol{x}_{t} \mid \boldsymbol{\theta}_\text{forecast}\right). 
\label{GP_mean_cov}
\end{align}
As described in the previous section, $m$ is the mean function of $f$, $K$ represents the covariance function of the GP, $\mathbb{I}$ represents an identity matrix, and $\sigma^2_\varepsilon$ denotes observation noise. Here, $\sigma^2_\varepsilon$ reflects aleatoric uncertainty (due to measurement errors), while $K$ reveals epistemic uncertainty (due to error in forecasting model).

\begin{table}[t]
\centering
\caption{Application Scenarios of Load and Renewable Energy Forecasting in Different Time Scales}
\small 
\begin{tabular}{p{3cm} p{4cm} p{9cm}} 
\toprule
\midrule
\textbf{Time Scale}\newline
    \textbf{(Forecast Range)} & \textbf{Application Scenarios} \\ 
\midrule
 \parbox{3cm}{Ultra-short-term\newline (Seconds to minutes)}& \begin{tabular}[c]{@{}l@{}}Power system frequency control \\ Real-time storage regulation\end{tabular} \\ \midrule
\parbox{3cm}{Short-term\newline (few hours to days)} & \begin{tabular}[c]{@{}l@{}}Day-ahead dispatch \\ Backup optimization\\ Unit commitment optimization \end{tabular} \\ \midrule
\parbox{3cm}{Medium-term\newline (One week to months)} & Maintenance plan \\ \midrule
\parbox{3cm}{Long-term\newline(Seasons or years)} & \begin{tabular}[c]{@{}l@{}}Substation planning and selection \\ Long-term generation plan \\ Design of renewable integration\end{tabular} \\\midrule
\bottomrule
\end{tabular}%
\label{forecasting}
\end{table}

\begin{figure}
  \centering
  \includegraphics[width=1\linewidth]{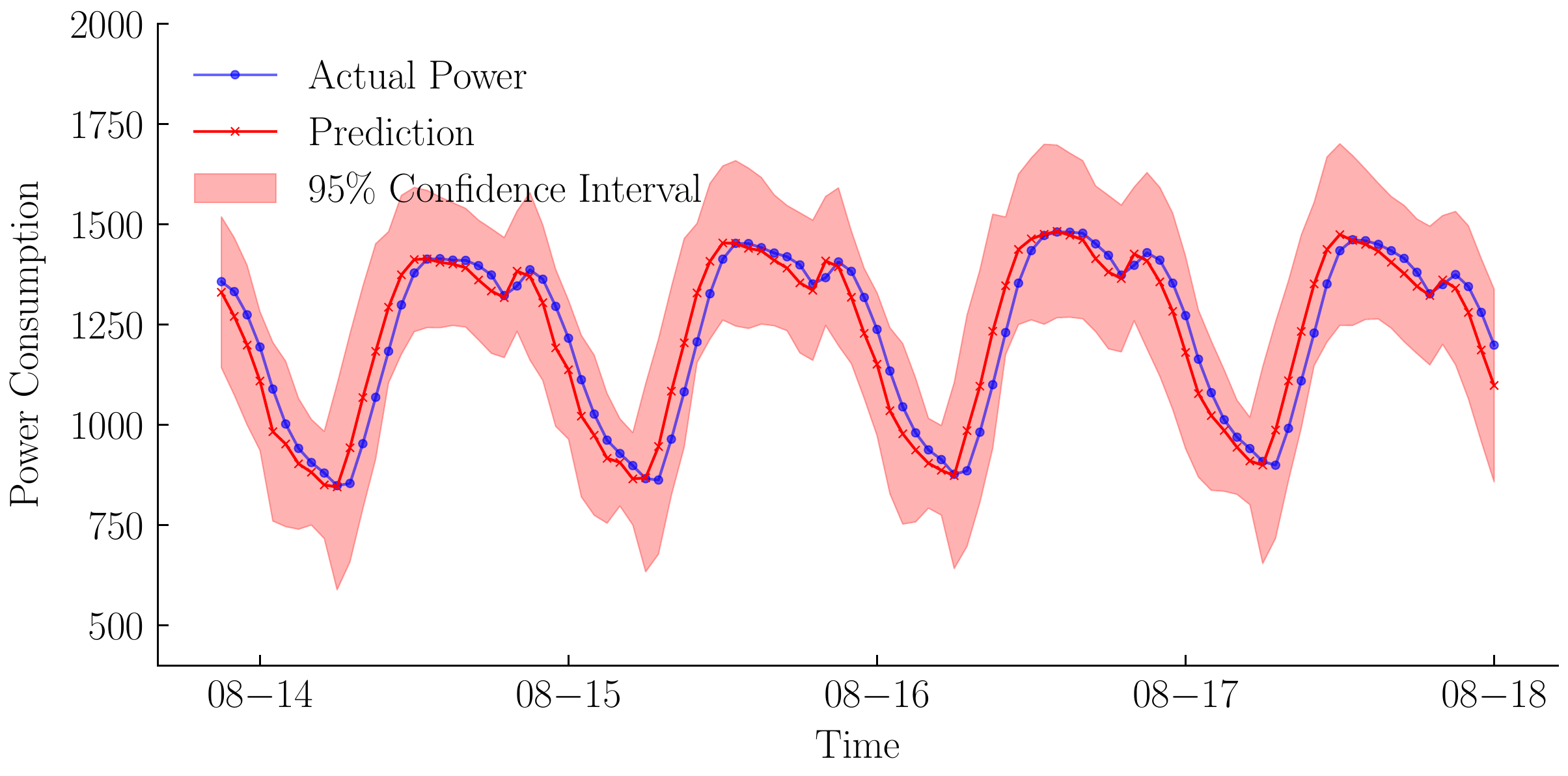}
  \caption{Load forecasting example using GP with squared exponential kernel.}
  \label{fig:load_forcasting_GP}
\end{figure}

An example of load forecasting using a GP model is illustrated in Fig.~\ref{fig:load_forcasting_GP}. The training dataset\footnote{Data can be obtained from \href{https://github.com/amirhosseinamerimanesh/Load-Forecasting-Machine-Learing}{here}.} comprises 12260 samples, each described by eight input features, including historical power consumption and temperature data with a 3-hour time lag. The training period spans from March 21, 2021, to August 13, 2022, covering diverse load consumption patterns. The GP model employs a squared exponential kernel with \(\nu^2 = 1\) and is trained over 50 iterations. For evaluation, data from August 13, 2022, to March 20, 2023, is used. Fig.~\ref{fig:load_forcasting_GP} specifically depicts forecasts for five days in August 2022. An extensive list of papers on enhanced GP-based forecasting tools for different time-scales as per Table \ref{forecasting}, is provided in Table \ref{forecasting_summary} along with the particular type of method used.

\begin{table*}[!t]
\vspace{-1em}
\centering
\caption{GP Applications in Power System Forecasting}
\footnotesize 
\begin{tabular}{>{\centering\arraybackslash}p{2cm} >{\centering\arraybackslash}p{2.8cm} >{\centering\arraybackslash}p{2.2cm} >{\centering\arraybackslash}p{3.5cm} >{\arraybackslash}m{5.5cm}} 
\toprule
\midrule
\textbf{Categories} & \textbf{Problems} & \textbf{References} & \textbf{Methodologies} & \textbf{Key Features} \\ 
\midrule
\multirow{8}{*}{\parbox{2cm} {\vspace{8.5\normalbaselineskip} \centering Forecasting feature engineering} } & \parbox{2.8cm} {\centering Short-term wind speed forecasting} & \cite{hu2017hybrid,hu2015short} & \parbox{3.5cm} {\centering Standard GP with Student-t observation model \cite{hu2017hybrid}, Standard GP \cite{hu2015short}}  & \multirow{2}{*}{\parbox{5.5cm} {\vspace{1\normalbaselineskip} Wavelet transform is employed to extract relevant features from the input data.} }  \\ \cmidrule{2-4}
& \parbox{2.8cm} {\centering Short-term solar radiation forecasting} & \cite{huang2016forecasting},\cite{ferkous2021wavelet} & \parbox{3.5cm} {\centering Standard GP} & 
  \\ \cmidrule{2-5}
& \parbox{2.8cm} {\centering Short-term PV power forecasting} & \cite{semero2018pv} & Standard GP & GP is integrated into a binary genetic algorithm framework to identify optimal input parameters.
 \\ \cmidrule{2-5}
& \parbox{2.8cm} {\centering Wind speed forecasting} & \cite{zhang2019wind} & Standard GP & Shared-weight LSTM is used to extract the features of input data. \\ \cmidrule{2-5}
& \parbox{2.8cm} {\centering Short-term wind power forecasting} & \cite{chen2013wind} & Censored GP & \multirow{4}{*}{\parbox{5.5cm} {\vspace{3\normalbaselineskip} ARD kernel is utilized to construct the covariance matrix of the GP, allowing for the calculation of each variable’s sensitivity during model learning.} } \\ \cmidrule{2-4}
& \parbox{2.8cm} {\centering Short-term load forecasting} & \cite{xie2018integrated} & Standard GP  & \\ 
\cmidrule{2-4}
& \parbox{2.8cm} {\centering Short-term load forecasting} & \cite{kou2014sparse} & \parbox{3 cm} {\centering Sparse heteroscedastic GP}  & \\ 
\cmidrule{2-4}
& \parbox{2.8cm} {\centering Short-term power consumption, PV power and net demand forecasting} & \cite{van2018probabilistic} & \parbox{3 cm} {\centering Standard GP}  & \\ 
\midrule

\multirow{8}{*}{\parbox{2cm} {\vspace{7.5\normalbaselineskip} \centering Forecasting model enhancement} } & \parbox{2.8cm} {\centering Short-term wind power forecasting} & \cite{wen2022sparse} & Sparse Variational GP & \multirow{4}{*}{ \parbox{5.5cm} {\vspace{1\normalbaselineskip} Variational sparse techniques leverage a reduced set of inducing points to approximate the full historical dataset, enabling efficient computation.}}  \\ \cmidrule{2-4}
& \parbox{2.8cm} {\centering Short-term load forecasting} & \cite{feng2024probabilistic}, \cite{ghasempour2023multiple} & \parbox{3.5cm} {\centering Sparse Variational GP \cite{feng2024probabilistic}, Multi-output sparse GP \cite{ghasempour2023multiple}} & \\ \cmidrule{2-4}
& \parbox{2.8cm} {\centering Short-term wind speed forecasting} & \cite{zhang2022novel}, \cite{wang2022sparse} & Sparse Variational GP & \\ \cmidrule{2-5}
& \parbox{2.8cm} {\centering Short-term wind speed forecasting} & \cite{zhang2016gaussian} & Standard GP & An autoregressive model is utilized as the mean function of the GP to capture the overall structure of the wind speed time series. \\ \cmidrule{2-5}
& \parbox{2.8cm} {\centering Short-term load forecasting} & \cite{dab2022compositional} & Standard GP & \multirow{2}{*}{ \parbox{5.5cm} {\vspace{0.3\normalbaselineskip} Composite covariance kernel functions are used to capture specific data characteristics, such as periodicity and local variability.}} \\ \cmidrule{2-4}
& \parbox{2.8cm} {\centering Short-term solar radiation forecasting} & \cite{lubbe2020evaluating} & Standard GP &  \\ \cmidrule{2-5}
& \parbox{2.8cm} {\centering Short-term load forecasting} & \parbox{2.2cm} {\centering \cite{cao2021robust}, \cite{li2023residential}, \cite{zhang2024multi}} & \parbox{3.5cm} {\centering Deep GP \cite{cao2021robust}, deep kernel learning \cite{li2023residential}, neural network GP \cite{zhang2024multi}} & End-to-end learning architectures, such as hierarchical GP structures or embedding neural networks into the kernel, are employed to enhance the expressiveness of GP.
 \\ \midrule
\multirow{6}{*}{\parbox{2cm} {\vspace{7.5\normalbaselineskip} \centering Forecasting output functionality} } & \parbox{2.8cm} {\centering Short-term load forecasting} & \cite{yang2018power} & \parbox{3.5cm} {\centering Quantile GP} & Deliver quantile predictions for load along with its associated probability density distribution \\ \cmidrule{2-5}
& \parbox{2.8cm} {\centering Short-term wind power forecasting} & \cite{rogers2020probabilistic} & \parbox{3.5cm} {\centering Heteroscedastic GP} & Automatically quantify time-varying noise in wind turbine power time series, using an
adaptive noise variance updating mechanism. \\ \cmidrule{2-5}
& \parbox{2.8cm} {\centering Short-term wind power forecasting} & \cite{heng2022probabilistic} & \parbox{3.5cm} {\centering Generalized GP} & Flexibly model wind speed uncertainty distributions. For instance, the likelihood function can be constructed using Weibull, truncated normal, or log-normal distributions. \\ \cmidrule{2-5}
& \parbox{2.8cm} {\centering Short-term load forecasting} & \cite{ghasempour2023multiple},\cite{gilanifar2019multitask} & \parbox{3.5cm} {\centering Multi-task GP} & \multirow{3}{*}{ \parbox{5.5cm} {\vspace{1.9\normalbaselineskip} Multi-task GP can be used to simultaneously forecast multiple time steps or integrate information from different geographical locations.}}\\ \cmidrule{2-4}
& \parbox{2.8cm} {\centering Short-term wind power forecasting} & \cite{liao2023probabilistic} & \parbox{3.5cm} {\centering Multi-task GP} & \\ \cmidrule{2-4}
& \parbox{2.8cm} {\centering Short-term wind speed forecasting} & \cite{cai2020gaussian} & \parbox{3.5cm} {\centering Multi-task GP} & \\ \midrule
\bottomrule
\end{tabular}%
\label{forecasting_summary}
\end{table*}
\subsection{Learning Steady-State Power Flow Solutions}
\label{sec:static_modeling}
In the power flow learning problem, the objective is to model the state of system (e.g., node voltage magnitude, branch flow) as a function of the load vector. In this general setting, one can consider the state as output while the load vector is input. Here, the load vector refers to the net load vector, obtained by considering the nodal generation as a negative load. Note that the relation between net load and voltages is given by the following power flow equations:
\begin{equation}\label{eq:pf}
\resizebox{.89\linewidth}{!}{$
\left\{
\begin{array}{l}
p_{g_{i}} - p_{d_{i}} = \sum\limits_{j\in\mathcal{N}} v_{i}v_{j} \left(g_{ij}cos(\theta_{ij}) + b_{ij}sin(\theta_{ij}) \right) \\
q_{g_{i}} - q_{d_{i}} = \sum\limits_{j\in\mathcal{N}} v_{i}v_{j} \left(g_{ij}sin(\theta_{ij}) - b_{ij}cos(\theta_{ij}) \right) ,
\end{array}
\right.$}
\end{equation}
where, $\mathcal{N}$ is the number of buses; $p_{g}$ and $q_{g}$ represent active and reactive power generations while $p_{d}$ and $q_{d}$ the demands, their difference being the negative net load; $g_{ij}$ and $b_{ij}$ are conductance and susceptance between bus $i,j$ respectively; $\theta_{ij} \!=\! \theta_{i} \!-\! \theta_{j}$ is the angle difference. However, the mapping from input (net load) to output (voltages), as described by \eqref{eq:pf}, is nonlinear and not easily invertible, making it computationally challenging. Here, we describe a standard GP-based probabilistic power flow model for the voltage function that can be evaluated efficiently. We then describe advances in kernel design for improved power flow learning. 

Let $f_s(\bm{x})$ represent the inverse power flow function from net load $\bm{x}$ to observed voltage measurement $\bm{y}$ at node $k$ as 
\begin{align}\label{eq:v_gp}
  y(\bm{x}) = f_s(\bm{x})+\varepsilon.
\end{align}
Here, $\varepsilon \sim \mathcal{N}(0, \sigma^2_\varepsilon)$ denoting independent Gaussian noise with zero mean and variance $\sigma^2_\varepsilon$. The node subscript $k$ is omitted for simplicity. Using the GP notation, $\bm y(\bm{x}) \sim \mathcal{GP}\big ( \bm{0}, K(x,x) +\sigma_\epsilon^2\mathbb{I} \big )$, 
where $K(\cdot,\cdot)$ represents the kernel matrix for $N$ training samples. 
Since voltage behaves smoothly as a function of load, the squared exponential kernel has been widely adopted for power flow learning tasks \cite{liu2022kernel, pareek2021framework, xugp1, xugp2}. The parameters of the model can be selected using the maximum likelihood estimator as described in the previous section. The detailed results in \cite{pareek2020optimal} show that GP-based approximations excel over traditional linearizations \cite{deka2017structure} of power flows under uncertainty. While the model uses a standard GP, we can improve its performance by selecting/designing a specific kernel based on power flow insights as described next.

\textbf{Kernel designs for power flow:}
Although the standard squared exponential kernel is most widely used for power flow approximations, they have limitations regarding scalability \cite{pareek2021framework}. It is worth noting that the maximum likelihood for the standard squared exponential kernel is done over a $2|\mathcal{B}|$ dimensional space (double of system size), and requires a high number of samples for accurate learning. A computationally efficient kernel design strategy is known as the vertex degree kernel (VDK), which can also be interpreted as physics-informed kernel design \cite{pareek2024fast}. 

The VDK kernel in \cite{pareek2024fast}, is based on the additive GP kernel idea, and constructed using sub-kernels, each defined as a \textit{squared exponential kernel} restricted to the neighborhood of a node. For node $j$, the sub-kernel operates on $\bm{x}_j = \{x_i | i = j \text{ or } (ij) \in \mathcal{E}\}$. This design leverages the insight that while a node's voltage depends on all loads, significant correlated effects are confined to nearby loads, making distant loads nearly uncorrelated. 

The VDK is a sum of these sub-kernels, expressed as:
\begin{align}\label{eq:VDK}
  k_{v}(\bm{x}^k, \bm{x}^l) = \sum_{b=1}^{|\mathcal{B}|} k_b(\bm{x}^k_b, \bm{x}^l_b).
\end{align}
Here, $k_b(\bm{x}^k_b, \bm{x}^l_b)$ is a sub-kernel operating on the subset of $\bm{x}_b$ voltages \footnote{See Fig. 2 in \cite{pareek2024fast} for graphical illustration of the VDK construction.}. Each sub-kernel $k_b$ has hyperparameters $\boldsymbol{\theta}_b$. The VDK GP has been extended to model network contingencies by changing its sub-kernels and also by adding additional kernels \cite{pareek2023data}. 

\subsection{Learning Power System Dynamics}
\label{sec:dynamic_modeling}
Unlike the steady-state power flow defined in Section \ref{sec:static_modeling}, we now describe the dynamic behavior over time of power systems with uncertain resources. The dynamic behavior can be described by the following system of differential and algebraic equations (DAEs) \cite{nguyenTAC, PAREEK2020106545}:
\begin{equation}
\label{eq:dae}
\begin{array}{l}
\left\{
\begin{array}{l}
\bm{\dot{y}} = \bm{\phi}(\bm{y},\bm{z},\bm{u},\bm{x}) \\
\bm{0} = \bm{\Psi}(\bm{y},\bm{z},\bm{x})
\end{array},
\right.
\end{array}
\end{equation}
where $\bm{\phi}(\cdot)$ and $\bm{\Psi}(\cdot)$ are vector-valued functions that describe the dynamics of synchronous machines, renewable energies, and loads, as well as the associated power flow equations. Here, $\bm{y}$ and $\bm{z}$ represent the dynamic state variable vector and the algebraic variable vector, respectively, encompassing variables such as bus voltage magnitudes and phase angles; $\bm{u}$ denotes system input; and $\bm{x}$ is a random vector representing uncertainties from dynamic loads and renewable energies, such as power injections and parameters. By utilizing the structural-preserve approach, \eqref{eq:dae} can be reformulated into the following differential equation \cite{Ye_2023}:
\begin{equation}
\label{eq:dae_Kron}
\bm{\dot{y}} = \bm{\phi}(\bm{y},\bm{\Psi}^{-1}(\bm{y},\bm{x}),\bm{u},\bm{x}),
\end{equation}
where the algebraic variables $\bm{z}$ are represented by $\bm{\Psi}^{-1}(\bm{y},\bm{x})$. This can be further linearized following the Euler-based explicit scheme:
\begin{equation}
\label{eq:dae_Kron_lnr}
\bm{y}(t) = \bm{y}(t-1) + \bm{\phi}\big(\bm{y}(t-1),\bm{\Psi}^{-1}(\bm{y}(t-1),\bm{x}),\bm{u},\bm{x}\big)\Delta t,
\end{equation}
where $\Delta t$ is the time step in the simulation process. 
To investigate the impact of uncertain resources on dynamic responses, the relationship between $\bm{x}$ and noise ($\varepsilon$) corrupted state variable measurements $\bm{y}$ is rewritten into the following compact form:
\begin{equation}
\label{eq:uq_model}
\bm{y}(t) = f_d(\bm{x},t)+\varepsilon,
\end{equation}
where $\bm{x}$ refers to uncertainties from loads and renewable energies, and $y$ contains relevant dynamic response variables. \eqref{eq:uq_model} can be regarded as an extension of \eqref{eq:v_gp} from a static to a dynamic context. We can model this relation using a GP with special kernels. Multi-task kernel might be useful in modeling the temporal patterns in the input and output relationship. \cite{Ye_2023} models \eqref{eq:uq_model} by constructing sparse GPs separately for each time step of the dynamics, while \cite{Ye_2024} further utilizes a shared kernel to develop a partially parallel GP, which is a multi-output GP model, to predict all system dynamics simultaneously.

\subsection{Probabilistic Optimal Power Flow}
\label{sec:POPF}
The probabilistic optimal power flow (POPF) problem provides statistical descriptions of generator setpoints and system states in optimal power flow solutions for a distribution of inputs\cite{pareek2020gaussian}. Unlike the standard OPF problem, the POPF problem accounts for uncertainty in inputs (typically load or renewable generation), with the objective of obtaining the output distributions.

POPF can be formulated as learning an input-output mapping $f: \; \bm{x} \mapsto [\bm{u^*};\bm{y}]$, where \(\bm{x} \in \mathbb{R}^n\) collects uncertain injections (e.g., renewable generation, real/reactive loads), $\bm{u^*}$ is the optimal generator set points and \(\bm{y} \in \mathbb{R}^m\) consists of system states ( node voltages, line flows etc.), that arise from solving the AC-OPF problem
\begin{align}
  \min_{\bm{u}} ~~ & c(\bm{u}) \label{POPF}\\
  \text{s.t.} ~~ & \bm{g}(\bm{x}, \bm{y},\bm{u}) = \bm{0}, \; \bm{h}(\bm{x}, \bm{y},\bm{u}) \leq \bm{0}. \nonumber
\end{align}
Here, $c(u)$ is the generation cost, \(g(\cdot)\) enforces power balance and \(h(\cdot)\) represents operational limits of voltage and line flows etc.

In the GP-POPF approach \cite{pareek2020gaussian}, the mapping \(f(\cdot)\) is modeled non-parametrically using GP regression:
$f(x) \sim \mathcal{GP}\left(0, k(x,x')\right)$, with a squared exponential kernel. After training on a dataset of \(N\) deterministic OPF samples \(\{(\bm{x}_i, (\bm{u^*}_i,\bm{y}_i))\}_{i=1}^N\), GP-POPF provides closed-form posterior means and variances for outputs, enabling uncertainty propagation under arbitrary input distributions without assumptions. Additionally, interpretability is introduced via the subspace-wise Sensitivity, defined as the ratio of kernel hyperparameters. 

In the next section, we will discuss the use of GP-based steady-state and dynamic models of power flow states presented in this section, for the problem of risk assessment, where the risk of violating stability or reliability conditions is estimated.

\section{Power System Risk Assessment}
\label{sec:risk}
The motivation to advance risk assessment methodologies in power systems is driven by the evolving complexities of modern grids and the integration of diverse, variable energy sources \cite{SHIWEN20171200}. In this context, there is a clear shift from traditional deterministic risk models to probabilistic, quantitative approaches. We first describe static and dynamic risk and then describe the use of GP to model both.

\textbf{Static risk assessment problem:}
In static power system risk assessment, the goal is to determine if power system variables $y$, such as bus voltage magnitudes, angles, and branch flows, will violate stability and reliability boundaries for the range of input conditions. We can formulate it as follows \cite{Tan2024Scalable}:
\begin{equation}
\label{eq:uq_model_static}
y = f(\bm{x}),
\end{equation}
where $f$ denotes the system model which can, for example, represent an AC power flow equation listed in \eqref{eq:pf}; $\bm{x}$ represents the vector of uncertain inputs, including factors like power injections from loads, PVs, or wind generation; Using the same framework as \eqref{eq:v_gp}, $y$ can be learned using training data by a GP framework, with estimated mean and covariance functions that define its distribution. For assessing risk, we can calculate the probability of exceeding a predefined line limit threshold \cite{10688568}:
\begin{equation}\label{eq:error_risk}
   \mathrm{Risk} = \mathbb{P}(\lvert Y \rvert > Y_{\text{lim}}) 
   = \int_{Y_{\text{lim}}}^{\infty} p_{|Y|}(y) \,{\rm d}y ,
\end{equation}
where, $Y_{\text{lim}}$ is the pre-defined threshold necessitated for state $y$; $p_{|Y|}(y)$ is the PDF.
Furthermore, we can compute the expected value of the excess over the threshold, i.e. the expected excess (EE) as an index for risk severity \cite{10688568}:
\begin{equation}\label{eq:error_excess}
\begin{aligned}
  {\rm Severity}
  &= E\big[\left\lvert Y \right\rvert - Y_{\text{lim}} \mid \left\lvert Y \right\rvert > Y_{\text{lim}}\big] \\
  &= \frac{\int_{Y_{\text{lim}}}^{\infty} (\left\lvert y \right\rvert - Y_{\text{lim}}) p_{\lvert Y \rvert}(y) \,{\rm d}y}{\mathbb{P}(\lvert Y \rvert > Y_{\text{lim}})}.
\end{aligned}
\end{equation}

In \cite{PAREEK2021132941,pareek2024fast}, GP is used for voltage injection mapping and to compute the risk of voltage violations. The application of GP in distribution networks derives closed-form expressions mapping load vectors to node voltages, effectively quantifying the probability of violation of voltage limits \cite{9552521}. For system-level risks, \cite{10688568} applies GP to probabilistic power flow analysis for overloading risk assessment under contingencies, while \cite{9281551,9052481} focuses on the use of GP for analyzing load margin relationships to address voltage violations. 

\textbf{Dynamic risk assessment problem:}
Dynamic risk involves the assessment of system behavior over time, especially during and after disturbances. Unlike static risk, dynamic risk must account for time-dependent interactions between synchronous machines, dynamic loads, and renewable generation sources under uncertain conditions and propagation of corresponding uncertainties. Building upon the dynamic power system model in \eqref{eq:uq_model}, GP has been used as an efficient surrogate model for dynamics for small-signal stability assessment in \cite{PAREEK2020106545,zhai2023online}.

Another way to use GP in dynamic risk assessment is to directly learn and predict metrics used in stability, such as transient stability index (TSI), region of attraction (ROA), and rate of change of frequency (ROCOF). While TSI assesses the ability of the power system to maintain synchronous operation following disturbances \cite{prob_stability_anal} and can be estimated using GP \cite{Ye_2023}, ROA defines the set of initial conditions under which the system will converge to a stable equilibrium point \cite{Tan2023Transferable, chaoonline}. ROCOF \cite{9693288, jalali2021inferring} quantifies how rapidly frequency changes can occur due to sudden load shifts or generation loss. Similarly, load margin is another crucial metric that indicates the maximum additional load that can be supported before the system reaches a state of instability \cite{Tan2023Transferable,Tan2022Interpretable} and can be modeled as a GP \cite{KANWAL2018865}. Further, for broader stability analysis, \cite{chaoROA} employs GP to estimate Lyapunov functions from voltage angle measurements, being among the first works to introduce GP in power system stability analysis. Furthermore, \cite{Tan_2025} proposes a stability pattern-informed multi-output GP to predict post-fault dynamic trajectories of rotor angle and speed.

\textbf{Extended works on risk assessment:} Aside from these methods for static and dynamic risk assessment, historical data can be directly used to predict failures in equipment using GP that relate to system risk. \cite{Hachino2015improvement} employs GP to predict potential damage to the power grid by forecasting the number of power outages. In microgrid protection applications, both \cite{SRIVASTAVA2022107889} and \cite{9625829} leverage GP for fault detection and location prediction. \cite{LI2019357} utilizes GP in a classification model for diagnosing wind turbine faults. \cite{app11010153} employs GP for early failure prediction in hydropower units, allowing effective monitoring and issuing early warnings. 

In the next section, we will discuss the use of GP-based models inside power grid optimization and control modules for enhanced computational speed-up. 

\begin{table*}[!t]
\centering
\caption{GP Applications in Power System Optimization and Control}
\footnotesize 
\begin{tabular}{>{\centering\arraybackslash}p{2cm} >{\centering\arraybackslash}p{2.8cm} >{\centering\arraybackslash}p{2.0cm} >{\centering\arraybackslash}p{3cm} >{\arraybackslash}m{6cm}} 
\toprule
\midrule
\textbf{Categories} & \textbf{Problems} & \textbf{References} & \textbf{Methodologies} & \textbf{Key Features} \\ 
\midrule
\multirow{5}{*}{\parbox{2cm}{\vspace{4.0\normalbaselineskip} \centering Static optimization and control for transmission systems}} 
& \parbox{2.8cm} {\centering Chance‑constrained OPF} & \cite{qin2024data} & Standard GP & GP regression is used to model uncertainty from load and renewable energy generation fluctuations, generating multiple uncertainty scenarios.
\\ \cmidrule{2-5}
& Stochastic AC-OPF & \cite{mitrovic2023data} & Standard GP & \multirow{2}{*}{\parbox{6.0cm} {\vspace{-0.05\normalbaselineskip} GP is used to approximate the AC power flow equations under input uncertainty and is then embedded as a chance-constrained surrogate model in the OPF.}}
\\ \cmidrule{2-4}
& \parbox{3cm} {\centering Chance‑constrained AC-OPF} & \cite{mitrovic2023fast} & Hybrid \& sparse GP & 
\\ \cmidrule{2-5}
& Probabilistic OPF & \cite{pareek2020gaussian} & Standard GP & GP can directly obtain the distribution of OPF output variables for a given input uncertainty, eliminating the need to repeatedly solve OPF problems during uncertainty analysis.
\\ \cmidrule{2-5}
& AC-OPF & \cite{canyasse2017supervised} & Standard GP & Supervised learning is used as a real-time proxy for AC-OPF by predicting OPF feasibility indicators and costs.
\\ \midrule

\multirow{5}{*}{\parbox{2cm}{\vspace{7.5\normalbaselineskip} \centering Static optimization and control for distribution systems}} 
& \parbox{2.8cm} {\centering Multi-period stochastic OPF} & \cite{bauer2023analytical} & Standard GP & Generation and storage schedules are optimized using a stochastic chance-constrained multi-period OPF, which accounts for time-varying uncertainties and incorporates a large energy storage system.
\\ \cmidrule{2-5}
& \parbox{2.8cm} {\centering Voltage regulation for active distribution system} & \cite{su2024analytic} & NNGP & The high-accuracy NNGP is reformulated as a chance constraint and analytically embedded into real-time voltage regulation, considering PVs, batteries, EVs, and topology changes.
\\ \cmidrule{2-5}
& Fast inverter control & \cite{jalali2022fast} & Standard GP & GP directly estimates inverter setpoints, enhanced by sensitivity information for accuracy and random features for efficiency, eliminating the need for OPF solving.
\\ \cmidrule{2-5}
& Volt-VAR control & \cite{olowu2023gaussian} & Standard GP & GP is used to predict the objective functions of the distribution OPF and is combined with a multi-objective genetic algorithm to determine the Pareto optimal solution.
\\ \cmidrule{2-5}
& \parbox{2cm} {\centering Optimal steady-state voltage control} & \cite{pareek2020optimal} & Standard GP & A linear voltage–power relationship for distribution system voltage control is constructed using GP, which is then used as a surrogate model for power flow calculations.
\\ \midrule

\multirow{2}{*}{\parbox{2cm}{\vspace{0.1\normalbaselineskip} \centering Static optimization and control for microgrids}} 
& \parbox{2cm} {\centering MPC for interconnected microgrids} & \cite{gan2020data} & \parbox{3cm} {\centering Co-regionalised GP} & \multirow{2}{*}{\parbox{6.0cm} {\vspace{0.1\normalbaselineskip} GP is used for time-series modeling to predict PV power generation and load demand, followed by MPC, which minimizes the objective function subject to constraints over the specified horizon.}}
\\ \cmidrule{2-4}
& \parbox{2cm} {\centering MPC for energy management system} & \cite{lee2018optimal} & \parbox{3cm} {\centering Standard GP} & 
\\ \midrule
\parbox{2cm} {\centering Building energy management} & \parbox{2.8cm} {\centering Demand response services with buildings} & \cite{nghiem2017data} & Standard GP & The building power demand response to control signals is modeled by a GP with confidence measures, enabling a model predictive controller to optimize building and battery control.
\\ \midrule

\multirow{3}{*}{\parbox{2cm}{\vspace{4\normalbaselineskip} \centering Dynamic optimization and control}} 
& \parbox{2.8cm} {\centering Transient stability preventive control} & \cite{su2023deep} & DSPP & DSPP is used to predict system transient stability with high computational efficiency while accurately quantifying the confidence intervals of the predictions, which is then embedded into the OPF to compute preventive control strategies.
\\ \cmidrule{2-5}
& \parbox{2.8cm} {\centering Transient stability emergency control} & \cite{su2025safe} & DSPP & DSPP is used for probabilistic transient stability estimation of the islanded microgrid while the safe deep reinforcement learning effectively solves the model while satisfying all hard constraints.
\\ \cmidrule{2-5}
& \parbox{2.8cm} {\centering MPC for single machine power system} & \cite{lee2018optimal} & Standard GP & Multiple GP models are trained to develop multi-step-ahead predictors for phase angles in the transient state, enabling MPC to optimize input signals by minimizing prediction errors relative to a reference.
\\ \midrule

\parbox{2cm} {\centering Unit commitment} & \parbox{2.8cm} {\centering Optimal unit commitment} & \cite{nikolaidis2021gaussian} & Standard GP & GP constructs a surrogate model of the unit commitment objective and uses an acquisition function for global exploration in areas of minimum mean and/or higher uncertainty.
\\ \midrule
\bottomrule
\end{tabular}%
\label{Table_Optimization}
\end{table*}
\section{Power System Optimization and Control}
\label{sec:optandcontrol}
GP-based methods are widely used in optimization and control problems within power systems, as well as in co-optimization problems involving other systems. As discussed in Section \ref{sec:POPF}, the map from input to solution of an optimization problem can be \emph{directly} modeled as a GP. In this section, we overview another approach, where GP-based models for non-linear system constraints (e.g., power flows constraints) or control laws are included \emph{within} a larger optimization problem as part of a learning-embedded optimization and control framework.

\textbf{Problem formulation:} A standard power system optimization and control problem consists of an objective function, along with equality and inequality constraints \cite{grigsby2007power}:
\begin{align}
\label{Eq_OPF}
\min ~~ & c(\bm{x},\bm{y}, \bm{u})\\
  \text{s.t.} ~~ & \bm{g}(\bm{x}, \bm{y},\bm{u}) = \bm{0},\;
\bm{h}(\bm{x}, \bm{y},\bm{u}) \leq \bm{0},\nonumber
\end{align}
where $\bm{y}$ represents the observed variables, $\bm{x}$ denotes the state variables; $\bm{u}$ denotes the control variables; $c(\bm{x},\bm{y}, \bm{u})$ is the objective function; $\bm{g}(\bm{x}, \bm{y},\bm{u})$ are equality constraints; $\bm{h}(\bm{x}, \bm{y},\bm{u})$ are inequality constraints. This is a generalization of the formulation in \eqref{POPF}. Following the discussion in the previous section, inequality constraints related to nodal voltages, battery operations, and stability constraints such as rotor angle and ROCOF can be incorporated into the constraints of \eqref{Eq_OPF} using GP-based models \cite{pareek2020gaussian}. Additionally, feedback control can be modeled by making input $\bm{u}$ as a function of changes in observation $\bm{y}$. We describe three specific forms of \eqref{Eq_OPF}: grid control, chance-constrained grid optimization, and co-optimization problems.

\subsection{Grid Optimal Control}
For control problems related to power and energy systems, GP provide differentiable, explicit forms of various physics-based equations. For example, in \cite{pareek2024degradation}, GP is used to model the battery degradation profile with respect to power dispatch, which is then used as a constraint in the overall battery dispatch problem. Furthermore, by leveraging the GP function, a sensitivity matrix that establishes the relationship between voltages and power injections is estimated, as discussed in Section \ref{sec:application}. This matrix can subsequently be used for eigenvalue, singular value or sensitivity analysis \cite{10688568,jalali2021inferring,jalali2022inferring,jalali2022fast}. 

The closed-form expression can be inserted into the control problem \eqref{Eq_OPF} in the form of \cite{pareek2020optimal,weng2022hypothesis}:
\begin{equation}\label{eq:OptCntr}
\bm{y}=\mathcal{M}\bm{x}+ \mathcal{L} \bm{u},
\end{equation}
where $\mathcal{M}$ and $\mathcal{L}$ represent the GP-based model for the dependence between observed $\bm{y}$ and state $\bm{x}$ and control $\bm{u}$, respectively. In addition, feedback control problems to compute the minimum and optimal control efforts, can be represented as \cite{pareek2020optimal}:
\begin{equation} 
\label{eq:OptCntrSS}
\bm{u} = \mathcal{M}^+ \Delta \bm{y},
\end{equation} 
where $^+$ denotes the pseudo inverse operator; control is designed based on change in $\bm{y}$. Integrating these derived closed-form solutions, problems such as voltage control, and battery sitting and sizing optimization \cite{pareek2021probabilistic} can be tackled. By taking advantage of GP-based power flow calculations, \cite{weng2022hypothesis} formulates a control problem aimed at minimizing the probability of feasibility violations. In \cite{nghiem2017data}, the response of the buildings to the control signal is modeled by a GP, which can predict the power demand of the buildings and improve the demand response performance of the system. In \cite{jalali2022fast}, a GP is used to predict the inverter control law in the presence of DERs in distribution grids.\cite{olowu2023gaussian} proposes a two-path constrained multi-objective framework to determine the optimal volt-VAR droop for grid-connected smart inverters. 

\subsection{Grid Optimization under Input Uncertainty} 
AC optimal power flow (AC-OPF) determines the cost-optimal generator set points to serve a given load vector, while satisfying the nonlinear AC power flow equations and system constraints. GP-based surrogates for the nonlinear power flow models can be included in OPF for computational speed-up. To ensure feasibility under input uncertainty, OPF solutions can be made uncertainty-aware by including generator recourse to the optimal generator set point 
\begin{align}
\bm{u} = \bm{u^*} + \bm{g}(\Delta \bm{x}).\nonumber
\end{align}
Here, the recourse $g(\Delta x)$ is a function of the input uncertainty $\Delta \bm{x}$. Further, the inequality constraints are replaced with their stochastic counterparts to model probabilistic constraint satisfaction under input uncertainty. One way of formulating this involves extending the OPF problem to a scenario-based framework with constraints pertaining to multiple input realizations \cite{vrakopoulou2013probabilistic}. Here, GP-based power flow models can be used to replace the non-linear power flow constraints for each load scenario. The second approach relies on converting probabilistic constraints into analytical chance constraints \cite{roald2017chance} to account for input uncertainties. As GP has closed-form gradients, they can be used to model approximate non-linear chance constraints by propagating uncertainty from input to output. In \cite{mitrovic2023data,mitrovic2023fast}, GP is used to approximate AC power flow equations, and then used inside chance constraints using Taylor approximation or exact moment matching of the GP equations. In \cite{qin2024data}, GP regression is used to model uncertainty from load and renewable energy generation fluctuations and then to generate multiple uncertainty scenarios that can be used inside a scenario-based OPF problem formulations. In \cite{su2024analytic}, an analytic neural network Gaussian process (NNGP)-based chance-constrained real-time voltage regulation method is proposed, where NNGP is used to construct voltage chance constraints. \cite{su2023deep} proposes a deep sigma point process (DSPP)-assisted chance-constrained power system transient stability preventive control method to address stability risks induced by uncertain renewable energy and loads. In \cite{hachino2015model}, multi-step-ahead predictors for the phase angle in the transient state of the power system are developed by GP and used inside a model predictive control (MPC) framework. 
\begin{table*}[!b]
\centering
\caption{Other Power System Applications of GPs}
\footnotesize 
\begin{tabular}{>{\centering\arraybackslash}p{2.5cm} >{\centering\arraybackslash}m{1.5cm} >{\centering\arraybackslash}m{1.5cm} >{\arraybackslash}m{10.8cm}} 
\toprule
\midrule
\textbf{Problems} & \textbf{References} & \textbf{Methodologies} & \textbf{Key Features} \\ 
\midrule
 \parbox{2.5cm}{\centering Transmission system state estimation}
  & \cite{konstantinou2021resilient} & Standard GP & The mean and covariance of state estimation obtained from GP are used as prior constraints in weighted least squares-based state estimation, reinforcing the resilience of system state estimation.  \\ 
\midrule
 \parbox{2.5cm}{ \centering Distribution system sensitivity analysis}& \cite{ye2021global} & Deep GP & Deep GP identifies the mapping relationship between power injections and voltages, facilitating the calculation of Sobol indices for power injections.  \\ 
\midrule
 Dynamic modeling   & \cite{mitrentsis2021probabilistic, mitrentsis2022gaussian} & Standard GP & GP maps the relationship between voltage and power at the interface of an active distribution system.
 \\
\midrule
 Signal recovery  & \cite{zimmer2024signal} & Multi-task GP & The multi-task kernel in GP captures the similarity among different signal channels, aiding the recovery of signals from other channels. \\
\midrule
\parbox{2.5cm}{ \centering Power system dynamics inference}   & \cite{jalali2021inferring, jalali2022inferring} & Standard GP & Missing or unmeasured sensor data is reconstructed by transforming a multi-input multi-output system into multiple single-input single-output systems, where inference is performed using GP within these systems. \\
\midrule
\bottomrule
\end{tabular}%
\label{Other applications}
\end{table*}
\subsection{Co-optimization Problems} Co-optimization problems often involve multiple objectives, sometimes with conflicting interests. They typically consist of multiple components or layers, resulting in high computational complexity. In cross-domain co-optimization, the problems often consist of systems with fundamentally different physical and operational characteristics. In such contexts, GP-learned surrogate models, which can accurately capture the relationships between system-specific variables and coupling variables, are especially well-suited, offering a flexible means to bridge distinct domains or sub-systems. In power systems' co-optimization problems, such as large-scale transmission network analysis and transmission–distribution co-optimization \cite{weng2022asymmetrically}, closed-form GP expressions for steady-state power flows are frequently employed as substitutes for AC power flow to reduce computational intensity while maintaining high accuracy across uncertainties. In cross-domain problems, GP-learned surrogate models not only contribute to reducing the computational complexity but also help bridge the modeling and information exchange between heterogeneous systems. The cross-domain modeling can be written as: 
\begin{equation} \label{eq:coop_GP}
\bm{r} = \hat{\bm{f}}(\bm{x},\bm{x}_{cp}),
\end{equation}
where $\bm{r}$ denote the selected output variables, $\bm{x}$ the set of input variables, and $\bm{x}_{cp}$ the set of coupling variables. $\hat{\bm{f}}$ represents the GP approximated function. The function \eqref{eq:coop_GP} can serve as a representative operational profile of a system, or a responsive function to changes in coupling variables. For example, the entire set of data center operational models is represented by a GP-leaned function during the co-optimization between power systems and data centers \cite{ liu2023gaussian,weng2023distributed}. Beyond computational benefits, function \eqref{eq:coop_GP} also acts as an insulation layer that enhances privacy and data security between different authorities, making them particularly valuable and practical in cross-domain co-optimization scenarios.

An extended list of applications of GP in power system optimization and control are summarized in Table \ref{Table_Optimization} in the Appendix.

\subsection{Other Applications}
In addition to power system forecasting, modeling, risk assessment, and optimization/control, GP has also been applied in distribution system state estimation (DSSE) \cite{cao2022topology, hu2024robust, dahale2021multi, cao2024forecasting, dahale2022recursive, zhang2022multi, zhang2023deep}. On one hand, to fully utilize the available data in the system for DSSE, measurements at different time scales \cite{dahale2021multi} and fidelities \cite{zhang2022multi, zhang2023deep} are integrated within the GP framework. On the other hand, uncertainties in pseudo-measurements and topology knowledge are incorporated into the GP kernel to enhance the robustness of DSSE against real-time measurement variations \cite{cao2024forecasting}. Additionally, multi-task GP is employed to capture correlations among different system topologies, enabling adaptivity to topology changes \cite{cao2022topology, dahale2021multi, dahale2022recursive} or facilitating multi-area estimation \cite{hu2024robust} in DSSE. There are also other interesting applications, such as distribution system sensitivity analysis \cite{ye2021global}, transmission system state estimation \cite{konstantinou2021resilient}, dynamic modeling for active distribution systems \cite{mitrentsis2021probabilistic, mitrentsis2022gaussian}, dynamic composite load modeling \cite{Tan_2025_load}, signal recovery \cite{zimmer2024signal} have also been successfully addressed using GP. A detailed summarization of these applications can be found in Table \ref{Other applications}.

\section{Enhancements and Future Works}\label{sec:challenge}
This section discusses the insight gained from experiences using GP in various applications in the power systems presented above, with general guidelines of when to use or not to use GP. Furthermore, we discuss some avenues for future research in GP, tailored for power systems applications. 

\textbf{Lesson learned:}
GP provides a probabilistic data-driven framework for uncertainty-aware modeling of input-output relationships. GP training can be efficiently done over limited data and its performance can be further improved by using active learning, where limited but useful training data is generated on the fly. Thus, GP is useful for power systems that warrant risk-aware decision making and computational efficiency. Note that several GP kernels are continuous functions that are analytically differentiable. GP performs favorably in applications that involve real-valued quantities for inputs and outputs, such as power grid states, generator state points, as well as control signals. This is particularly true for applications in estimation, modeling, and forecasting, as well as planning and operational problems that utilize the developed models. On the other hand, applications that require integer-valued outputs, such as fault localization, switching, unit commitment, etc., may not be the right application for GP.

\textbf{Future extensions:} While GP-based methods show promise in power grid operations, there are additional improvements desired to ensure their interpretability, scalability, robustness, online adaptiveness for power grids with a high number of heterogeneous complex devices and higher stochasticity. They are elaborated below.

\subsubsection{Interpretability} 
GP models allow a better understanding of the underlying function that is being learned. For instance, \cite{pareek2021framework} demonstrates how the hyperparameters of square exponential kernel can be used to assess the relative nonlinearity of various node voltage functions. However, this form of interpretability is largely qualitative and has limitations in providing quantitative or actionable insights. To mitigate this issue, several solutions can be proposed, such as integrating GP models with interpretable artificial intelligence techniques \cite{butler2024explainable}, designing kernel functions with clear physical interpretations \cite{chen2023identifiability, dai2020interpretable, pareek2024fast}, and embedding physical laws and power system constraints directly into the GP model \cite{Ye_2023}. For example, \cite{dahale2022recursive} designed a graph filter based on the graph Laplacian matrix, which is applied to the GP's covariance matrix. This is equivalent to encoding power system topology information into the kernel function. Furthermore, \cite{dab2022compositional,lubbe2020evaluating} employ composite covariance kernel functions to capture specific data characteristics, such as periodicity and local variability. 

\subsubsection{Feature engineering} This refers to the selection of appropriate input features and historical window to enhance model accuracy. In an unsupervised manner, \cite{hu2017hybrid,hu2015short,huang2016forecasting,ferkous2021wavelet} utilize wavelet transform to extract useful features from the inputs of GP, while also mitigating noise in these inputs. Some studies employ deep learning-based feature extractors, such as shared-weight LSTM \cite{zhang2019wind}, which are subsequently used as inputs to GP, significantly reducing the number of variables. However, with the increasing integration of highly nonlinear power electronic devices into the grid, the traditional two-stage data-driven modeling approach, namely feature extractors followed by GP, may be insufficient to capture the complex characteristics of power systems, particularly in dynamic analysis scenarios. This limitation stems from the nature of GP basis functions, such as the widely used Gaussian kernel, which offer smooth interpolation but struggle to generalize across a wide range of extrapolations~\cite{wilson2016deep}. To address this issue, end-to-end architectures such as Deep GP~\cite {su2023probabilistic} and deep neural networks (DNN)-kernelized GP~\cite{Tan2024Scalable} have been proposed to enhance the generalization capability of GP by extracting more expressive representations directly from the data. However, introducing such automated feature extraction either significantly increases the computational complexity of Bayesian inference or demands substantially larger datasets due to the increased number of trainable parameters. More works are required to enhance the feature embedding capabilities of GP models.

\subsubsection{Enhancing output functionality} This refers to providing additional information (beyond the mean and variance of a single GP target output. In this context, \cite{yang2018power} proposes quantile GP regression, which delivers quantile predictions for load along with its associated probability density distribution. Meanwhile, \cite{rogers2020probabilistic} introduces heteroscedastic GP to automatically quantify time-varying noise in wind turbine power time series, using an adaptive noise variance updating mechanism. Furthermore, \cite{chen2013wind,xie2018integrated,kou2014sparse,van2018probabilistic} apply the automatic relevance determination (ARD) kernel to construct the covariance matrix of the GP, allowing for the calculation of each variable's sensitivity during model learning. Recent studies~\cite{ghasempour2023multiple, liao2023probabilistic, gilanifar2019multitask, cai2020gaussian} have extended GP to a multi-output setting, enabling simultaneous prediction of vector-valued outputs relevant to power grids. However, as the output dimension increases, the computational complexity grows cubically. Therefore, future research could focus on improving computational efficiency in multi-output settings.

\subsubsection{Scalability} In standard GP, the computational complexity is \( \mathcal{O}(N^{3}) \), primarily due to the inversion of the kernel covariance matrix. This complexity limits GP's capability to handle large datasets (e.g., \( N > 10000 \). For single-output cases, sparse approximation reduces GP complexity by using \( M \) inducing points to approximate the dataset, lowering computational complexity to \( \mathcal{O}(NM^{2}) \) \cite{seeger2004gaussian}. Furthermore, stochastic variational inference can reduce this further \cite{hensman2013gaussian}. However, these approximation methods often sacrifice model accuracy. Future research can focus on enhancing GP scalability while preserving model accuracy, which remains a crucial area for power system analysis. For instance, at the data requirement level, physical model information from power systems can be incorporated to design the GP structure, thereby reducing the amount of data needed to train the GP. Recent progress in graphics processing unit (GPU)-accelerated computation also enhances the scalability of GP. Libraries such as GPyTorch \cite{gardner2018gpytorch} leverage parallelized matrix operations and efficient linear solvers to enable scalable training and inference on modern GPU hardware. These developments make it scalable to apply GP in power system applications. One possible solution is to integrate the robust statistics theory into the GP hyperparameter estimation process, achieving robust model parameter estimation even in the presence of non-Gaussian noise and outliers. 

\subsubsection{Robustness} In GP models, observations are assumed to be noisy realizations of an unknown latent function, which follows a Gaussian prior. To obtain the posterior distribution of the latent function, the noise is also assumed to follow a Gaussian distribution governed by the “nugget" parameter. While this assumption makes the posterior distribution tractable, it can render GP less robust. However, in practical power system operations, measurement errors may not strictly follow a Gaussian distribution \cite{zhao2018robust}; even when they do, their variance may be time-varying \cite{rogers2020probabilistic}. Additionally, power system data inherently contains outliers, which deviate significantly from the expected data range and may result from inaccurate measurements or cyber-attacks \cite{chakhchoukh2019diagnosis}. To enhance the robustness of GP against these measurement errors, a common approach is to model measurement error as a mixture of Gaussian distributions or to approximate it with a heavy-tailed distribution, such as Student’s t-likelihood function \cite{algikar2023robustgaussianprocessregression}. To effectively reduce the impact of outliers in time-series stochastic power flow, \cite{algikar2023robust} employs the Huber loss function and projection statistics to constrain the mean function. Existing methods do not generalize well to scenarios involving arbitrary measurement error distributions. Therefore, future research can focus on generalizing the likelihood function to ensure both a tractable GP posterior and robustness to diverse error distributions and extreme values.

\subsubsection{Online adaptiveness} This refers to the ability of GP to perform online learning over temporal or spatial data. Power systems are inherently dynamic, with sensors continuously collecting time-varying data. As mentioned earlier, training GP with cumulatively growing datasets becomes computationally infeasible under streaming conditions due to the cubic complexity \( \mathcal{O}(N^{3}) \). To support real-time tracking of system behavior, various online GP frameworks have been proposed so that GP can be updated with only a few newly obtained data points. For instance, \cite{zhai2022dynamic,zhai2021estimating} utilize online GP to estimate the region of attraction for operating states in real time, while \cite{kou2013sparse} applies online GP to wind power forecasting. Nevertheless, most existing methods lack mechanisms for real-time kernel adaptation or for handling structural changes in system topology. Addressing these limitations remains a promising research direction. To this end, \cite{lu2022incremental} proposes an ensemble GP approach that adaptively adjusts kernel weights for improved incremental learning on streaming data. Building on this, \cite{polyzos2021ensemble} introduces a Bayesian framework for the online adaptation of ensemble GP under evolving graph structures. Future work could further integrate power system-specific characteristics, such as network constraints, into online GP models to enhance their adaptiveness under time-varying operating conditions.

\subsubsection{Integration with other data-driven methods}
Deep learning models, such as DNN, can learn meaningful representations from high-dimensional data through multiple layers of highly adaptive basis functions \cite{wilson2016stochastic}. GP has been integrated with other methodologies, particularly DNNs, to enhance their performance. In voltage stability analysis, \cite{Tan2023Transferable} proposed an innovative combination of GP kernels with DNN to replace traditional continuation power flow methods for load margin assessment, effectively handling uncertainties in wind generation and load conditions. Building upon this work, \cite{Tan2024Scalable} developed a DNN-kernelized vector-valued GP to assess rare events risks, particularly focusing on voltage instability and power flow failure prediction. In wind power applications, \cite{MANOBEL20181015} combined GP with artificial neural networks to address uncertainty in turbine power prediction to improve data quality. \cite{LIO2021670} combined GP with the extended Kalman filter to estimate real-time rotor effective wind speed, effectively reducing risks associated with inaccurate wind speed measurements and enhancing generation stability. In transient stability analysis, \cite{li2014transient} integrated GP with kernel principal component analysis to develop a binary classification framework for power system transient stability assessment. Moreover, integrating GP with reinforcement learning can further extend their applicability and optimize performance in system control tasks \cite {su2024review, su2025safe, cai2021safe, lim2020prediction}. 
Furthermore, integrating GP with federated learning can enable communication-efficient global modeling across distributed devices in the system \cite{achituve2021personalized}. In addition, embedding GP within a nonlinear autoregressive framework can enhance system identification performance for nonlinear dynamic systems, such as inverter-penetrated power systems \cite{sarkka2021use}. These integrated approaches demonstrate how GP's probabilistic modeling capabilities can be enhanced through combination with specialized techniques. However, such combinations may compromise the stability of the models, especially under non-stationary or rare-event scenarios.

\section{Conclusion}\label{sec:conclusion}
This paper presents a comprehensive review of GP methodologies and their applications in power system operation and control. GP-based techniques have been extensively reviewed across several critical domains, illustrating their role in uncertainty quantification and decision support. While GP offers significant advantages in handling uncertainty in power system operation and control, further advancements are needed to enhance scalability, expressiveness, and robustness to enable real-world deployment in large-scale power grid analytics.

\bibliographystyle{IEEEtran}
\bibliography{reference.bib}



\end{document}